\documentclass[preprint2]{aastex}
\usepackage{natbib}

\begin{document}

\title{Seeing the sky through Hubble's eye: The COSMOS SkyWalker}

\author{K.~Jahnke}
\affil{Max-Planck-Institut f\"ur Astronomie, K\"onigstuhl 17, D-69117 Heidelberg, Germany}
\email{jahnke@mpia.de}

\author{S.~F.~S\'anchez}
\affil{Centro Astron\'omico Hispano-Alem\'an de Calar Alto, Calle Jesus
  Durban Rem\'on 2.2, E-04004 Almer\'ia, Spain}
\email{sanchez@caha.es}

\author{A.~Koekemoer}
\affil{Space Telescope Science Institute, 3700 San Martin Drive,
  Baltimore, MD 21218, USA}
\email{koekemoe@stsci.edu}

\begin{abstract}
 Large, high­resolution space­based imaging surveys produce a volume of
 data that is difficult to present to the public in a comprehensible
 way. While megapixel­sized images can still be printed out or downloaded
 via the World Wide Web, this is no longer feasible for images with 10$^9$
 pixels (e.g., the {\em Hubble Space Telescope} Advanced Camera for Surveys
 [ACS] images of the Galaxy Evolution from Morphology and SEDs [GEMS] project)
 or even 10 10 pixels (for the ACS Cosmic Evolution Survey [COSMOS]). We
 present a Web­based utility called the COSMOS SkyWalker that allows
 viewing of the huge ACS image data set, even through slow Internet
 connections. Using standard HTML and JavaScript, the application successively
 loads only those portions of the image at a time that are currently being
 viewed on the screen. The user can move within the image by using the mouse
 or interacting with an overview image. Using an astrometrically registered
 image for the COSMOS SkyWalker allows the display of calibrated world
 coordinates for use in science. The SkyWalker ``technique'' can be applied to
 other data sets. This requires some customization, notably the slicing up of
 a data set into small (e.g., 256$^2$ pixel) subimages. An advantage of the
 SkyWalker is the use of standard Web browser components; thus, it requires no
 installation of any software and can therefore be viewed by anyone across
 many operating systems.
\end{abstract}


\section{Introduction}
In the last several years, a number of large, space­based extragalactic
imaging surveys have been conducted with the {\em Hubble Space Telescope}
(HST) and its Advanced Camera for Surveys (ACS). Very prominent among these
are contiguous areas covered by the Hubble Ultra Deep Field
(UDF),\footnote{See the Hubble Ultra Deep Field project home page,
http://www.stsci.edu/ hst/udf.} the GOODS \citep[Great Observatories Deep
Origins Survey;][]{giav03} field, that of the GEMS \citep[Galaxy Evolution
from Morphologies and SEDs;][]{rix04} and STAGES (Space Telescope A901/902
Galaxy Evolution Survey; M. Gray et al. 2006, in preparation) projects, or by
COSMOS \citep[`Cosmic evolution survey';][]{scov06}, here ordered by
increasing sky coverage. While the UDF covers one ACS pointing, or about
3\arcsec~$\times$~3\arcsec, GOODS consists of 15 pointings, GEMS 80, STAGES
81, and COSMOS 575, with a total area of 2~deg$^2$ for the latter. All of
these images have in common that they are deep (the UDF is 1--1.5 mag deeper
than the Hubble Deep Field; the other surveys collected at least one HST orbit
per pointing), and they are very rich in galaxies: nearly 10,000 in the UDF,
40,000 in GEMS, and 2,000,000 in COSMOS.

These surveys all produced a very large number of data pixels, with up to
10$^{10}$ pixels for COSMOS alone. The image areas are contiguous, and for
parts of the surveys, they even exist as color images. So they are in
principle excellent material not only for science, but also for presentation
to the public.  However, their size makes it almost impossible to show both
size and detail at the same time. As an example, the UDF looks very nice
printed at a size of, say, 1.5~$\times$~1.5~m (corresponding to a resolution
of $\sim$5 pixels mm$^{-1}$), whereas the COSMOS ACS image would similarly
require at least 35~m on a side for a print of the same resolution. At the
same time it needs to be viewed from a distance of less than 1~m in order to
appreciate all the details. This is not practically feasible.

The presentation of such data to the public via the Internet is similarly
difficult. Even in a compressed JPEG format, the complete COSMOS ACS image
cannot be transported in a sensible time, and even current 64~bit PCs have
relevant size constraints for loading large images. In order to allow viewers
from the public or science communities to browse through such available data
sets, we have put together a World Wide Web-based application we call
``SkyWalker'', which allows the user to, in principle, browse arbitrarily
large images and to view any part of them on the screen. It builds on HTML and
JavaScript only, which are by default integrated in any current
image­capable Web browser. In this way, such images can be viewed not only
without the installation of specialized software, but also by anyone with an
Internet connection (even a slow one).

\section{The SkyWalker technique}
The goal of the SkyWalker is to allow browsing of large images without the
requirement of downloading them as a whole. It should be possible with any
graphical Web browser, and certainly for Firefox/Mozilla and Internet
Explorer. The user should be able to pan around in the full image.

\begin{figure}[t]
\includegraphics[width=8cm]{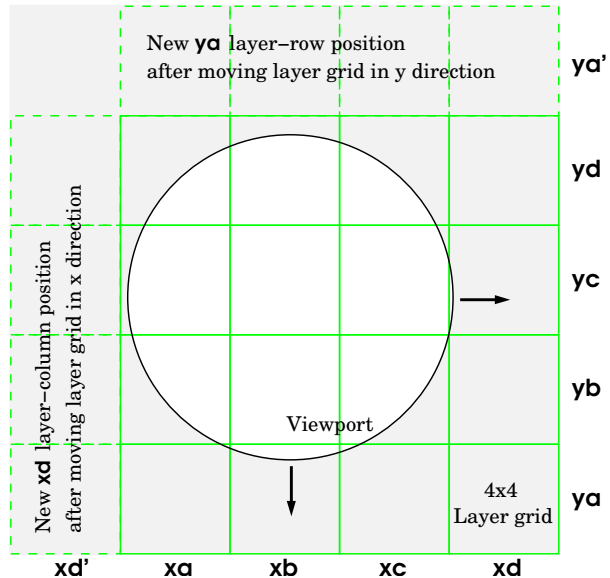}
\caption{
Layout of the SkyWalker image layer structure. A 4$\times$4 grid of layers
(columns marked xa,...,xd, rows ya,...,yd) contains the subimages (solid
squares) that connect smoothly at their borders. Only the area of the image
inside the viewport (circle) is visible. If one row or column of the layer
grid shifts outside the viewport, it gets relocated to the other side of the
grid (column xd$^\prime$ or ya$^\prime$; dashed squares), and the
corresponding images get loaded into the layers.
\label{fig:layout}
}
\end{figure}

The main technical solution is that the full image is sliced up into smaller
square images (``subimages'') of 256$\times$256 pixels, which are named with a
prefix and a running number.  At any time, 4$\times$4 adjacent subimages are
placed as background images in a grid of layers (``sublayers'') in the
framework of the JavaScript document object model \citep[DOM;][]{DOM}. While a
part of this region is visible through a mask (``viewport'') of $\le$768
pixels on a side, the remainder lies outside the visible region
(Fig.~\ref{fig:layout}).

In order to select the region to be viewed, the user either moves a pointer
around in a separate downscaled overview image that shows the location and
size of the viewport, or else directly drags the displayed image itself in
order to shift the position.  

All the JavaScript code required for this is contained in the SkyWalker HTML
page, which consists primarily of JavaScript and relies on the freely
available DynAPI library of cross­browser JavaScript
functions.\footnote{The DynAPI 2 Cross­Browser JavaScript Library,
Ver. 2.5.7, is available from http://dynapi.sourceforge.net under the GNU
General Public License.} The library bundles a number of common JavaScript
tasks, such as the creation, movement, and placement of layers, which are
applications that are used for the SkyWalker. By using DynAPI library
functions, a single syntax can be used for layer manipulations across
different browsers, which would otherwise vary.  All layer creation and
manipulation tasks are handled via this library.

From the positioning inside the overview image, which is also contained in its
own DOM layer, a JavaScript function computes several things: (1) the
associated pixel position of the full­scale image, (2) from the pixel
position the running numbers and thus the names of the 16 images to be
displayed, and (3) the screen coordinates of the 16 sublayers. These are
organized in four rows and four columns, and by always shifting
them to identical coordinates and by identical offsets, they appear as a
single, continuous image.  

If the image is shifted to a position such that, e.g., a column comes to lie
so far to the side that its contents are no longer visible in the viewport,
its layer position is relocated from the right to the left side of the
sublayer grid (or vice versa), and the image contents get exchanged to
continue the contiguous image on that side. In this way, the same 16 layers
are constantly defined and do not need to be destroyed and recreated again;
just their location and content is changed.

\begin{figure*}[t]
\centering\includegraphics[height=10cm]{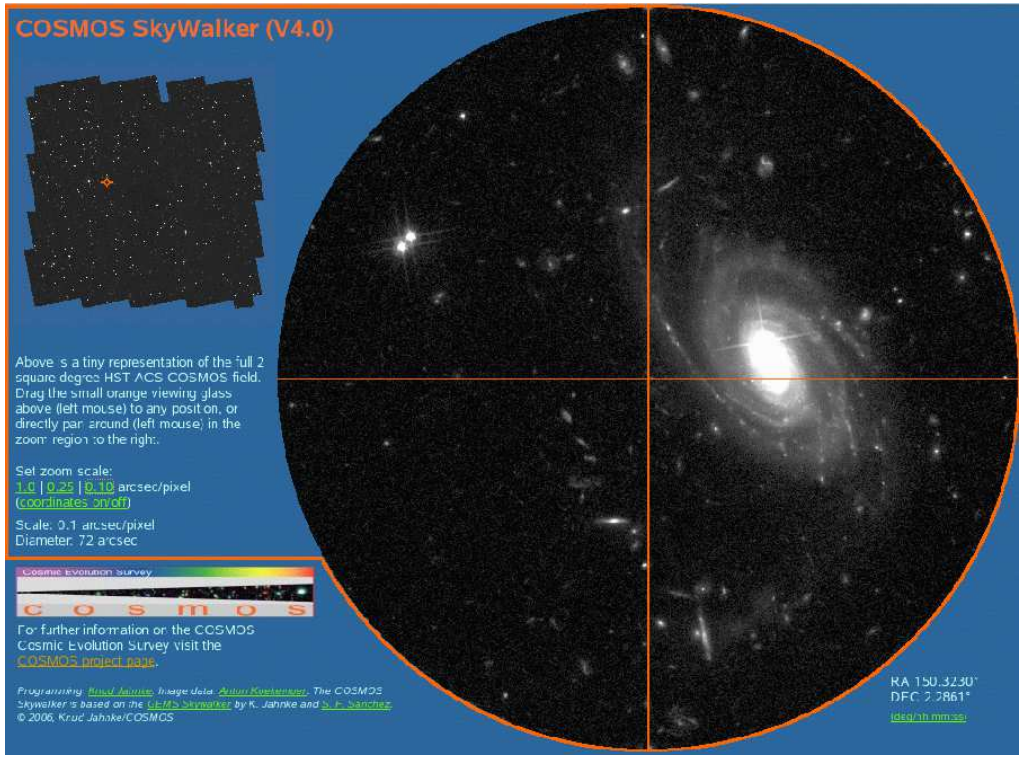}
\caption{
Screen shot of the COSMOS SkyWalker, showing the overview image (top left)
with the position crosshair in the main viewport (right), the coordinates
(bottom right) scale selection and image scale information (middle left). The
crosshair shows the coordinate position center, and it can be turned on or off
together with the coordinates.
\label{fig:swcosmos}
}
\end{figure*}

For the astrometrically calibrated COSMOS SkyWalker (Fig.~\ref{fig:swcosmos}),
three ``zoom levels'' are presented, consisting of sets of images of the
COSMOS ACS F814W images \citep{scov06b,koek06} at scale sizes of 1\farcs0,
0\farcs25, and 0\farcs1 pixel$^{-1}$. Other data sets are currently being
incorporated, notably from the VLA COSMOS radio survey; in the future, these
may potentially include data from other instru­ ments (e.g., X­ray and
other optical images). We also added the option of displaying the sky
coordinates of the central pixel in the viewport when activating a crosshair,
computing R.A. and decl. from the pixel coordinates in tangential projection.

Other features are conceivable for the future; e.g., accessing a catalog by
using information on individual galaxies and objects. Although the COSMOS
SkyWalker is not able nor is it meant to replace applications such as Skycat,
DS9, or other scientific data viewers, it can certainly be used for quick
visual access to the ACS data and for presentation of the material to
nonastronomers. The COSMOS SkyWalker is currently directly accessible at
\url{http://www.mpia.de/COSMOS/skywalker}.

\section{Future SkyWalking}
Given the current state of the code, it is in principle possible for the
SkyWalker application to present any contiguous data set that does not cover
more than one hemisphere.\footnote{4 Here we have to note that the current
version of the COSMOS SkyWalker works with images in gnomonic projection,
which, like all zenithal perspective projections, will lead to field
distortions at larger distances from the central point of the projection
\citep{cala02}. If fields larger than a few degrees are to be shown,
alternative projections should be used that are appropriate for the geometry
of the data set. To accommodate the full celestial sphere, any projection can
be used with nondiverging coordinates, the exact choice depending on whether
the projection is supposed to be, e.g., equiareal or equidistant. The
corresponding conversion from pixel to world coordinates then needs to be
added to the code.} In principle, it is possible to present any data set in a
SkyWalker application. However, several steps need to be taken in order to do
this.  As a prerequisite, the full image needs to be sliced up into subimages,
and a scaled­down overview image needs to be prepared. Translation
constants between the overview image and the full image have to be computed.

The SkyWalker code is open­source licensed under the GNU General Public
License (GPL) and is freely available from
\url{http://skywalker.sourceforge.net} (the COSMOS SkyWalker described here
represents ver. 4.0). At this site, we also maintain a list of different
SkyWalker applications, which currently point to SkyWalkers for UDF and GEMS
\citep{jahn04e} and COSMOS. We explicitly invite anyone to adapt current
versions of the SkyWalker to their own data set and unique needs, or to add
new features.

\acknowledgments
For the realization of the SkyWalkers, we would like to thank the COSMOS,
GEMS, and UDF teams for their work, which generated these images. We also
thank the developers of the DynAPI JavaScript library, which made programming
easy, and to the developers of STIFF at Terapix, which facilitated the
conversion of COSMOS ACS images from FITS to TIFF.

\bibliographystyle{apj}

\end{document}